# Night-to-Night Variability of Sleep Electroencephalography-Based Brain Age Measurements

Jacob Hogan[*1,2], Haoqi Sun[*1], Luis Paixao[1,3], Mike Westmeijer[1], Pooja Sikka[1], Jing Jin[1], Ryan Tesh[1], Madalena Cardoso[1], Sydney S. Cash[1], Oluwaseun Akeju[4], Robert Thomas[5], M. Brandon Westover[1]

[1] Department of Neurology, Massachusetts General Hospital, Boston, MA, USA

[2] Present Address: Department of Biology, Brigham Young University, Provo, UT, USA

[3] Present Address: Department of Neurology, Washington University School of Medicine in St. Louis, St. Louis, MO, USA

[4] Department of Anesthesia, Critical Care, and Pain Medicine, Massachusetts General Hospital, Boston, MA, USA

[5] Division of Pulmonary, Critical Care and Sleep Medicine, Department of Medicine, Beth Israel Deaconess Medical Center, Boston, MA, USA

* Contributed equally as co-first authors.

Corresponding Author:

M. Brandon Westover, M.D. Ph.D.

Department of Neurology, Massachusetts General Hospital

55 Fruit Street, Boston, MA 02114

*Abstract*

*Objective*
Brain Age Index (BAI), calculated from sleep electroencephalography (EEG), has been proposed as a biomarker of brain health. This study quantifies night-to-night variability of BAI and establishes probability thresholds for inferring underlying brain pathology based on a patient's BAI.

*Methods*
86 patients with multiple nights of consecutive EEG recordings were selected from Epilepsy Monitoring Unit patients whose EEGs reported as being within normal limits. BAI was calculated for each 12-hour segment of patient data using a previously described algorithm, and night-to-night variability in BAI was measured.

*Results*
The within-patient night-to-night standard deviation in BAI was 7.5 years. Estimates of BAI derived by averaging over 2, 3, and 4 nights had standard deviations of 4.7, 3.7, and 3.0 years, respectively.

*Conclusions*
Averaging BAI over $n$ nights reduces night-to-night variability of BAI by a factor of $\sqrt{n}$, rendering BAI more suitable as a biomarker of brain health at the individual level.

*Significance*
With increasing ease of EEG acquisition including wearable technology, BAI has the potential to track brain health and detect deviations from normal physiologic function. In a clinical setting, BAI could be used to identify patients who should undergo further investigation or monitoring.

*1. Introduction*

There are several dimensions of measurable brain health, including subjective complaints, neuropsychological testing (Lezak 2012), cerebral oxygen extraction and blood flow (Xu et al. 2009), brain structure using magnetic resonance imaging (Carne et al. 2006), and electroencephalographic (EEG) markers (Al Zoubi et al. 2018). The latter may be measured during wake or sleep. Wake characteristics of abnormal brain function include pathological slowing and abnormalities of alpha wave frequency and distribution. Sleep markers of brain health include spindle and delta power and K-complexes characteristics, which reflect health of the < 1 Hz non-rapid eye movement (NREM) sleep slow oscillation (Urakami et al. 2012). However, in isolation these biomarkers have shown only weak associations with disease state, and only at the group level.

The brain age index (BAI) is an algorithm for measuring how much an individual's electroencephalographic (EEG) brain activity during sleep differs from the expected pattern for an individual's age, based on a large number of features extracted from different sleep stages and the awake state. BAI is derived by first producing an estimate of the individual's age from an overnight sleep EEG, called the brain age (BA), and subtracting from this the individual's chronologic age (CA). Prior research suggests that, at the population level, brain age correlates strongly with chronologic age, and excess brain age (BAI>0) is associated with conditions that worsen brain function (e.g., diabetes, hypertension, dementia, psychiatric disease) (Sun et al. 2019). Thus, BAI holds promise as a biomarker of brain health.

However, before BAI can be applied in individual patients, it is necessary to understand its night-to-night variability, or its converse, night-to-night stability. Nightly variation in sleep EEG patterns may arise from either short-term factors not strongly related to long-term brain health, such as light or noise, body position, recent sleep loss, or other sleep disturbances (Brunner et al. 1993). By contrast, "brain health", as influenced by long-term exposure to factors such as cerebrovascular or neurogenerative disease, chronic intermittent hypoxia from sleep apnea, inflammatory states, exercise, nutrition, or long-term sleep habits, is expected to vary more slowly, perhaps over the course of weeks to months or even years, but to be relatively stable on a night-to-night time scale. If BAI shows high night-to-night variability, then it may be advantageous to measure BAI several nights in a row and average the measurements, to derive a meaningful biomarker of brain health more applicable at the level of individuals. By this reasoning, the degree to which a high BAI value should be considered "abnormal" depends on the extent to which it exceeds expected levels of night-to-night fluctuation.

In this study, we therefore aimed to: (1) determine the night-to-night variability of the brain-age index in individual patients, and (2) establish thresholds for the probability of underlying brain pathology based on a patient's BAI, i.e. values of BAI above which concern and further investigation are warranted.

*2. Methods*

**2.1. Patient Selection Criteria**

For this study, we obtained data from adult patients who underwent multi-day EEG recordings during the course of routine clinical care but were not acutely ill. The Epilepsy Monitoring Unit (EMU) provides such a cohort.  Patients admitted to the EMU include a mixture of those with established epilepsy (excluded from the present study), and those being evaluated to determine whether or not they have epilepsy. A portion of patients in the latter group ultimately prove to have normal EEGs; some of these are patients with epilepsy whose burden of seizures and epileptiform activity is low, while others are ultimately found to have problems unrelated to epilepsy (e.g., behavioral / psychogenic events or cardiac events masquerading as seizures). EEG data from these patients provide a unique opportunity to use existing clinical data to estimate night-to-night variability of EEG-based brain age in EEGs that are free of epileptiform abnormalities which would otherwise interfere with calculation of BAI.

We therefore used the following criteria to select patient data for our present study: eligible subjects included any adult patient (≥18 years) who underwent continuous EEG (cEEG) monitoring at the Massachusetts General Hospital Epilepsy Monitoring Unit (EMU) between May 1, 2014, and May 1, 2019, who had (1) at least two consecutive nights of EEG recordings; (2) the EEG recordings were "normal" (i.e., showed no signs of epileptiform EEG activity, as such abnormalities would affect the calculation of brain age) as determined by the clinical report signed by the attending clinical neurophysiologist; and (3) at least five hours of cEEG data for each 12-hour period. EEG outside of sleep times was not considered in this analysis. Five hours was chosen because it typically suffices to sample multiple epochs of each stage of sleep.

While no EEGs containing epileptiform activity were included, some patients had a diagnosis of epilepsy. Out of 1,147 patients admitted to the EMU during the 5-year period, eighty-six patients met the inclusion criteria. A patient selection flowchart is shown in Figure 1.

The Partners Institutional Review Board approved this retrospective analysis without requiring additional consent for its use in this study.

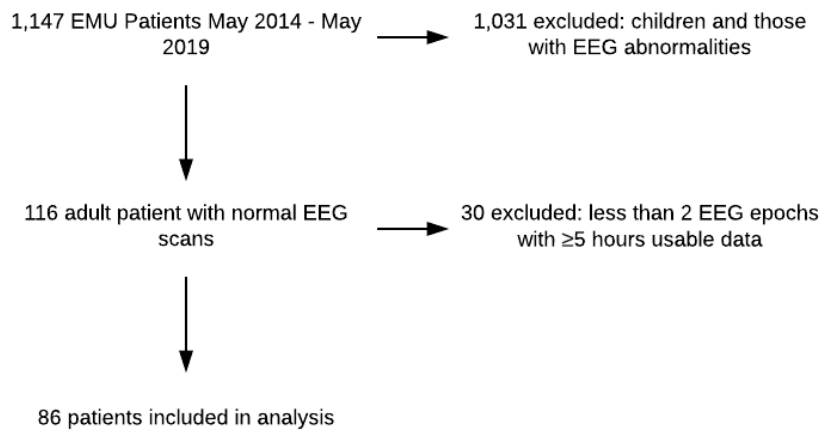

Figure 1- Patient Selection Flowchart.

## 2.2. Patient Cohort

Previous MRI studies show that epileptic patients' brains appear to show structural signs of accelerated biological aging (Pardoe et al. 2017; Sone et al. 2019). To test whether this observation also applies to BAI, we examined whether BAI differs in patients with vs without epilepsy. For this comparison we used a one-sided, two-sample Welch's t-test, with the hypothesis that epileptic patients would have a higher average BAI than those without epilepsy.

In addition, we examined the difference in within-patient standard deviation in BAI between epileptic patients and non-epileptic patients using a two-sided, two-sample t-test. This was done to verify our assumption that the standard deviation in BAI estimates does not vary significantly between the two groups.

### 2.3. EEG Preprocessing and Artifact Removal

Our study used 6 channels for sleep-staging: frontal (F3 and F4), central (C3 and C4), and occipital (O1 and O2). The EEG signals were notch-filtered at 60Hz and then bandpass filtered from 0.5 – 20Hz. EEG signals were divided into 30-second epochs for analysis. Each 30-second epoch was marked as artifact and removed if the epoch contained two or more seconds of saturated signal (amplitude greater than 500μV), or five or more seconds of flat EEG signal (standard deviation of the amplitude less than .2 μV for five seconds, or less than 1 μV for the whole 30-second epoch).

### 2.4. Sleep Staging

To classify EEG epochs into sleep-stages, a previously published extreme learning machine (ELM) based model, trained on a dataset of 2,000 PSGs, was used (Sun et al. 2017). This algorithm classifies each 30-second EEG epoch into one of five stages: awake, NREM1, NREM2, NREM3, or REM.

### 2.5. Removing Eye-open Awake Signals

The ELM based brain age algorithm was trained on data from patients in a sleep lab setting, consisting primarily of eyes-closed, in-bed periods, whereas the EMU EEG data used in the present study also includes prolonged periods when patients are active and awake. To render the data appropriate for the BAI algorithm, eyes-open awake epochs were removed using an automated process. This process uses a customized blink detection algorithm, as follows: After sleep staging, ten randomly-selected, one-minute awake EEG segments were hand labeled for blinks, and a parameter search was performed to optimize a threshold for peak detection (Gramfort et al. 2013, 2014), used in a blink-detection algorithm developed by MNE (a signal-processing software suite named for its ability to calculate Minimum Norm Current Estimates) (Gramfort et al. 2013). This algorithm first filters the signal using a Finite Impulse Response (FIR) and a band pass filter and then detects peaks. We used bipolar montage channels Fp1-F3 and Fp2-F4 channels in the blink detection routine. MNE source code can be found here (Larson 2019a, 2019b).The optimal threshold was 90μV, with a Cohen's kappa value of 0.83.

To remove eyes-open awake data, 30-second epochs which (1) contained greater than six blinks and (2) were classified as awake by the ELM based sleep-staging model were removed. On average, the eyes-open awake period comprised 15.1% of the data. After removing eyes-open awake epochs, patients with at least one epoch containing five hours of usable data were included. The mean number of available 5-hour epochs per patient during their EMU stay was 5.6.

### 2.6. Brain Age Night-to-Night Variability Analysis

EEG features and sleep-stages are required inputs for calculating BAI. Features were calculated using previously published methods (Sun et al. 2019). The previous study used frontal (F3-M2 and F4-M1), central (C3-M2 and C4-M1) and occipital (O1-M2 and O2-M1), with each channel referenced to the contralateral mastoid. Our study used the same 6 channels: frontal (F3 and F4), central (C3 and C4) and occipital (O1 and O2), except that the reference channel was C2 (second cervical vertebra) rather than the contralateral mastoid.

Our goal here was to quantify variability of BAI for a patient that has had $n$ close-in-time sleep EEGs (spaced apart by no more than a few days), hence $n$ BAI estimates. We propose both a "theory" and an empirical approach to verify the theory:

*2.6.1. Signal + Noise theory of multi-night BAI variability:*

We hypothesized that BAI measurements taken within a few days of each other can be modeled as a constant, the "true" underlying value of the BAI, plus independent noise:

$$\widehat{BAI} = BAI + \epsilon \text{ (Equation 1)}$$

We assume that the noise is normally distributed with zero mean, $\epsilon \sim N(0, \sigma^2)$, where $\sigma^2$ is the variance, to be estimated. Under this simple model, when more than one BAI measurement is available $\widehat{BAI}_1, \widehat{BAI}_2, \ldots, \widehat{BAI}_n$, we average the measurements to obtain an estimate with a higher signal to noise ratio, $\langle BAI_n \rangle = \frac{1}{n}\sum_{i=1}^{n} \widehat{BAI}_i$. Under this model, the standard error of the mean $\langle BAI_n \rangle$ decreases as $1/\sqrt{n}$, thus

$$\langle BAI_n \rangle = BAI + \epsilon', \text{where } \epsilon' \sim N\left(0, \frac{\sigma^2}{n}\right) \text{ (Equation 2)}.$$

Our primary objective is to estimate the standard deviation of BAI measurements, $\sigma$. We note that the assumption that the data follow a normal distribution is not strictly required for the variance to decrease as proposed. Nevertheless, we make this assumption for simplicity and subsequently verify that it provides a reasonable empirical fit to the data.

*2.6.2. Empirical approach to multi-night BAI variability:*

We take an empirical approach to verifying the theoretical model previously described. For explanatory purposes, we choose an example patient who has 7 independent measurements of BAI from 7 nights of EEG data (Figure 2). Since the objective is to characterize how averaging multiple independent BAI estimates reduces within-patient standard deviation in BAI, we calculate the within-patient standard deviation of BAI measurements that are averaged over $n$ nights. For the case where $n$ =1, a single BAI estimate is computed for each night (no averaging). In cases where $n$ is greater than 1, BAI estimates are calculated by averaging $n$ independent BAI calculations from $n$ nights.

In the case where $n$ =1 (no averaging), we calculate the standard deviation in brain age estimates for each patient using all available BAI values for each patient. For an example patient with seven nights of data, this would simply be the standard deviation of these seven estimates (Figure 2a).

Next, consider the case where BAI estimates from multiple nights are averaged together to arrive at a less noisy BAI estimate. We explain our approach beginning with the case where $n$ = 2 (Figure 2b). To estimate the standard deviation of averages based on pairs of BAI measurements, we partition each patient's single-night brain age estimates into groups of two. We use the word partition deliberately: to ensure that our estimates of variability are not downwardly biased, we do not take all possible pairs of two; rather, we allow any single observation to show up in only one pair of measurements. For a patient with seven brain age estimates, we calculate 3 means of BA estimates using 3 pairs: 1-2, 3-4, and 5-6, with brain age estimate 7 not used (Figure 2b). The within-patient standard deviation of these pair-based ($n$ =2) average BAIs is the standard deviation of these three independent values. Note that we do not consider all pairs of size 2 (of which there are 21) because many of those pairs (18 out of 21) would have an observation which shows up in at least one other pair, which would cause the mean calculated from those 2 pairs to be correlated and bias the estimated variability downward.

When $n$ =3, we partition the sample into groups of 3 independent BAI measurements (Figure 2c). We then calculate the average BAI for each of these groups. For our example patient, BAI estimates would be separated into two groups—containing estimates 1,2, and 3; and estimates 4,5, and 6—and groupwise average BAIs would be calculated. The within-patient standard deviation for this patient when $n$ =3 is the standard deviation between the two average BAI estimates (Figure 2c).

For a patient with seven brain age estimates, note that we could not calculate the standard deviation of BAI if $n$ =4 nights, because there aren't enough estimates to calculate two independent means of BAI which use 4 nights each. Thus, the standard deviation for estimates of brain age obtained by averaging 4 (or $n$) nights is only calculated if patients had at least 8 ($2n$) nights of EEG data (Figure 2d).

The stated procedure—partitioning each patient's brain age estimates into groups of size $n$, taking the average BAI for each group, and calculating the standard deviation of group averages—is repeated 1,000 times per patient using random permutations, and the average within-patient standard deviation in BAI is used for each patient.

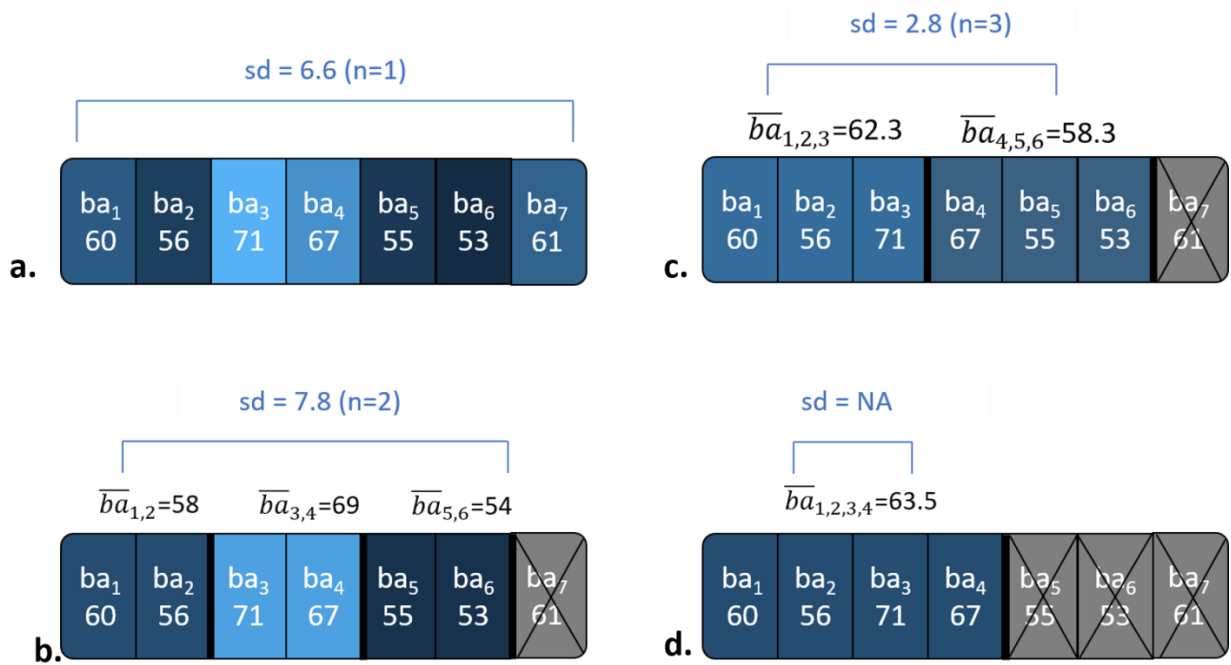

Figure 2- Diagram showing how the standard deviation in brain age index was calculated for $n$ = 1, 2, 3, and 4 nights for a single patient. For $n$ = 2-4, each procedure was repeated using 1,000 random permutations, and the results were averaged.

## 2.7. Effect of Sleep Stage distribution on BAI Variability

To understand whether BAI variability is related to the percent of time spent in each sleep stage, we first calculate the stage distribution (percent of time spent in each sleep stage) for each patient. Then, a regression model is fit, attempting to predict the patient's standard deviation in BAI from the percent of time spent in each stage using both univariate regression for each sleep stage, and multivariate linear regression using all sleep stages.

We also investigate whether BAI variability might be explained in part by night-to-night variability in a patient's sleep stage distribution. To test this possibility, we calculate the standard deviation in the percent of time spent in each sleep stage for each patient. Then, another regression model is fit, attempting to predict the patient's standard deviation in BAI from the standard deviation in percent of time spent in each stage using both univariate regression for each sleep stage, and multivariate linear regression using all sleep stages.

## 2.8. Defining Brain Age Risk Lookup Table

Using the mean intra-patient standard deviation for $n$ =1, 2, 3, and 4 nights, we calculate the probability that a patient's true brain age index (the BAI estimate based on an infinite number of nights) would be equal to their BAI estimate. Adopting a Bayesian framework, we assume that, conditional on the patient's true BAI, the distribution of brain age estimates ($\widehat{BAI}$) is randomly distributed around that true BAI, according to Equation 1 ($\widehat{BAI} = BAI + \epsilon, \ \epsilon \sim N(0, \sigma^2)$). As for the prior distribution over values of BAI, we assume it is uniformly distributed on an interval $[L, U]$. This choice of prior avoids any strong assumptions regarding whether patients undergoing BAI testing are from a healthy vs. sick population, in line with our goal of using BAI as a biomarker of brain health across a wide range of patients. Given this assumption, we apply Bayes rule to make inferences about a patient's actual brain age index, $BAI$, given their brain age estimate, $\widehat{BAI}$:

$$P(BAI|\widehat{BAI}) = \frac{P(\widehat{BAI}|BAI)P(BAI)}{P(\widehat{BAI})} \ \text{(Equation 2)}$$

Under the assumption that the distribution of brain age estimates—$P(\widehat{BAI}|BAI)$—is normally distributed around a patient's true BAI and the true distribution of patient BAIs—$P(BAI)$—is uniformly distributed, the posterior distribution $P(BAI|\widehat{BAI})$ is (approximately) a normal distribution. Thus, the probability distribution of a patient's true BAI is normally distributed around an unbiased estimate of BAI. Note that the posterior distribution is technically a truncated normal distribution; however, to reflect a lack of strong conviction regarding the true BAI's distribution, choosing a large interval $[L, U]$ for the uniform distribution approximately results in a normal distribution.

Using the normal distribution, the average standard deviation in brain age index estimates was then used to calculate the probability that a patient's brain age index is equal to a certain value $y$, given that their estimate of brain age index is equal to $x$. These results are reported as a probability density table. To facilitate usability within real world clinical and research contexts, we also calculated the probability that a patient's true underlying brain age is greater than or equal to a certain value $y$, given that their brain age estimate is equal to $x$. These results are presented in the brain age risk lookup table.

## 3. Results

### 3.1. Patient Cohort

Patient characteristics are shown in Table 1. Sleep stage summary values are found in Table 2.

Our patient cohort was 65% female ($n$=56), 73% ($n$=63) of patients had a diagnosis of epilepsy and 61% ($n$ =52) used anti-epileptic drugs during the EMU stay (Table 1). The cohort had a mean of 5.6 epochs of data per patient, a mean age of 42.0 (SD=13.8), and a mean BAI of 12.3 (SD=12.7). Total recording time was 5,472 hours, with a sleep efficiency of 0.73 (Table 2).

Although average BAI for epileptic patients (BAI=13.2, $n$=63) was higher than for non-epileptic patients (BAI=9.9, $n$=23) in our cohort, this difference was not statistically significant (p value = 0.14, one-sided two-sample Welch's t-test). Night-to-night variability (BAI standard deviation) was also not significantly different between these two groups (two-sided two-sample t-test, epileptic patients' BAI standard deviation average sd=7.7 vs. non-epileptic patients BAI average sd = 7.7; p = 0.49).

Table 1. Patient characteristics, medical history, and AED administration.

| **Patient Characteristics** | |
|---|---|
| n | 86 |
| Age (mean (SD)) | 41.97 (13.84) |
| Mean BAI (mean (SD)) | 12.30 (12.68) |
| Female Gender (%) | 56 (65.1) |
| White Race (%) | 81 (94.2) |
| **Medical History** | |
| Epilepsy (%) | 63 (73.3) |
| Hypertension (%) | 35 (40.7) |
| Tobacco (%) | 20 (23.3) |
| Diabetes (%) | 8 ( 9.3) |
| Substance use (%) | 4 ( 4.7) |
| **AED Administration** | |
| Any AED (%) | 55 (64.0) |
| Levetiracetam (%) | 19 (22.1) |
| Lamotrigine (%) | 19 (22.1) |
| Clonazepam (%) | 14 (16.3) |
| Gabapentin (%) | 16 (18.6) |
| Lorazepam (%) | 9 (10.5) |
| Other (%) | 10 (11.6) |

Table 2. Summary statistics of sleep.

| Sleep Summary | |
|---|---|
| Total Recording Time (hrs) | 5472 |
| Total Sleep Time (hrs) | 3984 |
| Sleep Efficiency | 0.73 |
| n epochs (mean (SD)) | 5.6 (3.4) |
| Epoch length (mean (SD)) | 11.4 (1.5) |
| **Sleep Stage Totals (hrs (%))** | |
| N1 | 599 (10.9) |
| N2 | 2,307 (42.2) |
| N3 | 101 (1.8) |
| REM | 978 (17.9) |
| Awake | 1036 (18.9) |
| nan | 453 (8.3) |
| **Patient Sleep Stage (hrs (%))** | |
| N1 | 1.2 (11.0) |
| N2 | 4.7 (41.8) |
| N3 | 0.2 (1.9) |
| REM | 2.1 (18.4) |
| Awake | 2.1 (18.6) |
| nan | .62 (13.0) |

## 3.2. Brain Age

Figure 3 shows two brain age estimates for two different patients. Patient 1 (2A and B) has similar brain age estimates for both nights, while patient 2 (2C and D) has different brain age estimates for multiple nights. Some aspects of night-to-night variability in sleep patterns can be seen in the EEG spectrogram shown in this figure.

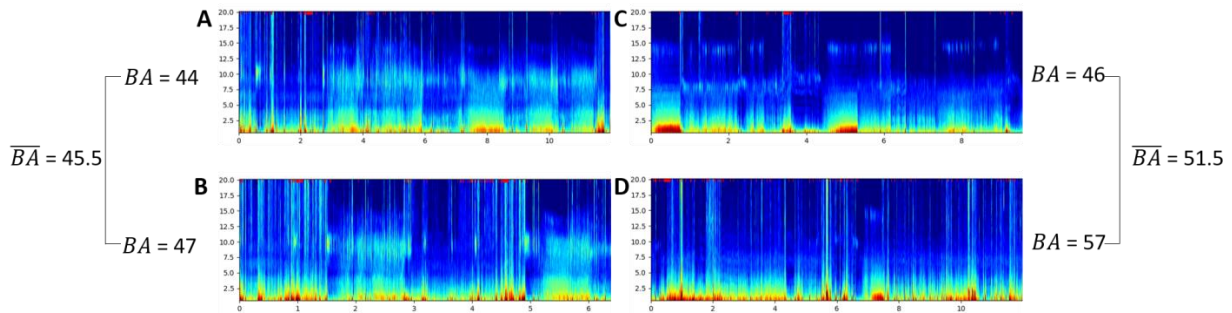

Figure 3- Example patient epochs and calculated brain ages (BA) for two patients. Patient 1 (left) has a stable BA estimate across two nights. Patient 2 (right) has two different BA estimates, with regions of increased delta power associated with a younger BA. Each subplot is a power spectrum for the patient for brain waves of frequency 0-60. Dark blue indicates low power, while red indicates high power.

### 3.3. Night-to-Night Variability

The estimated within-patient night-to-night standard deviation in BAI was 7.5 years. Estimates of BAI derived by averaging over 2, 3, and 4 nights, had lower estimated standard deviations of 4.7, 3.7, and 3.0 years, respectively. The standard error of the average of $n$ BAI estimates approximately decreases according to $1/\sqrt{n}$ as $n$ increases, as shown in Figure 4, in support of our "signal + noise" model of BAI night-to-night variability.

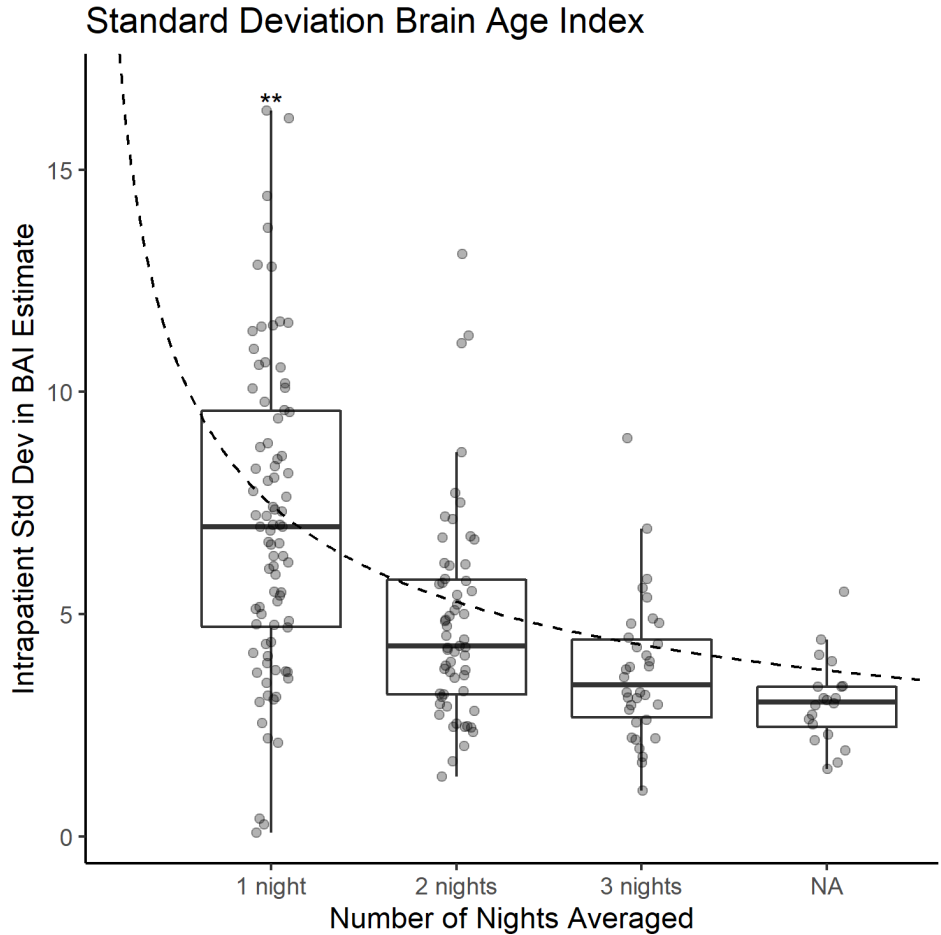

Figure 4- Average standard deviation in BAI for all patients when samples of sizes 1, 2, 3, and 4 nights are taken respectively. X axis: number of nights averaged, or $n$. The dashed line represents the decrease in standard error we would expect with increasing sample size (SD/$\sqrt{n}$). ** represents two outliers with BAI = 20 and 30 respectively.

### 3.4. Effect of Sleep Stages and Patient Covariates on BAI Variability

Neither the percent of time spent in each sleep stage nor the variability in percent of time spent in each sleep stage were significant predictors of night-to-night variability of BAI for any of the variables tested. P-values for univariate and multivariate linear regression are provided in Supplementary Tables S1 and S2. This suggests that variation in stage distribution does not significantly contribute to night-to-night variability of BAI.

**3.5. Brain Age Risk Lookup Table**

The results showing the probability that a patient's actual brain age is equal to $y$ given that their estimated brain age is $x$ are shown in the probability density tables of Figure 5. When $n = 1$, we see that the probability distribution is relatively flat and diffuse, reflecting relatively high night-to-night variability of BAI measurements for a single individual from a single night of sleep. By contrast, for a patient whose BAI is averaged over four nights, the probability that the true BAI is equal to their estimated BAI (within 0.5 years) is 12.9%, and further calculations show that there is a 50% probability that a patient's average brain age estimate from four nights is within 2.0 years of their actual brain age. We also see that the post-measurement probability distribution becomes decidedly more concentrated around the estimated value, reflecting the higher signal-to-noise ratio of multi-night BAI estimates.

The brain age risk lookup tables are also shown in Figure 6. These are similar to the probability density tables, except that probabilities shown are the likelihood that a patient's true brain age is greater than or equal to $y$ given an estimated brain age of $x$. Since increased brain age is viewed as concerning for the presence of pathology, this table can be used to calculate a patient's risk of having an abnormally high BAI.

The blue boxed value in Figure 6c can be interpreted as follows: A typical patient with an estimated BAI of 8 when averaging three nights of BAI estimates has a 91% chance that their true BAI is greater than or equal to 3. Other values can be interpreted similarly.

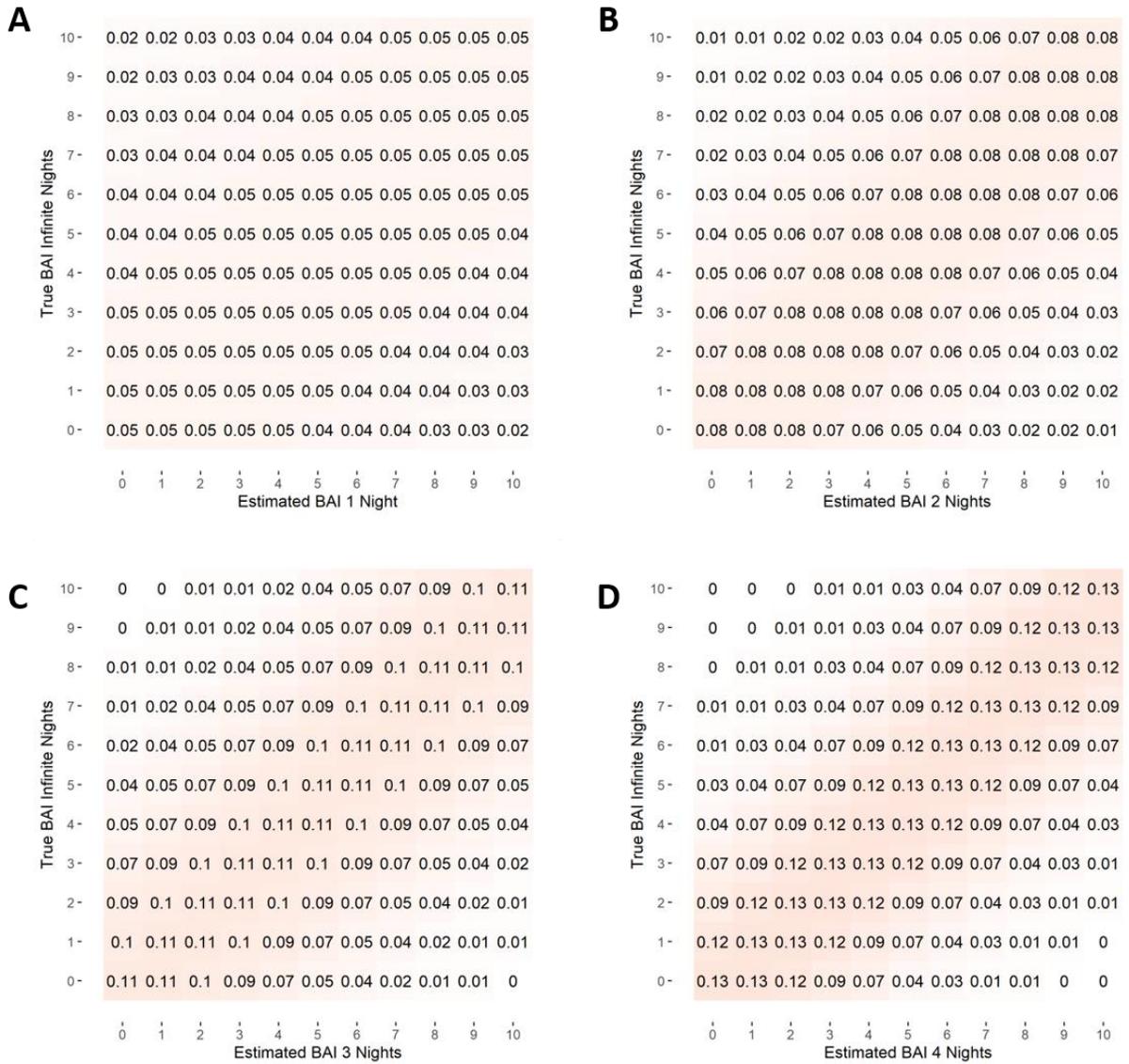

Figure 5- Table of probability density. The probability that a patient' true BAI is equal to (±.5years) their estimated BAI if the estimate is calculated from a single night, or the average BAI 2, 3 or 4 nights.

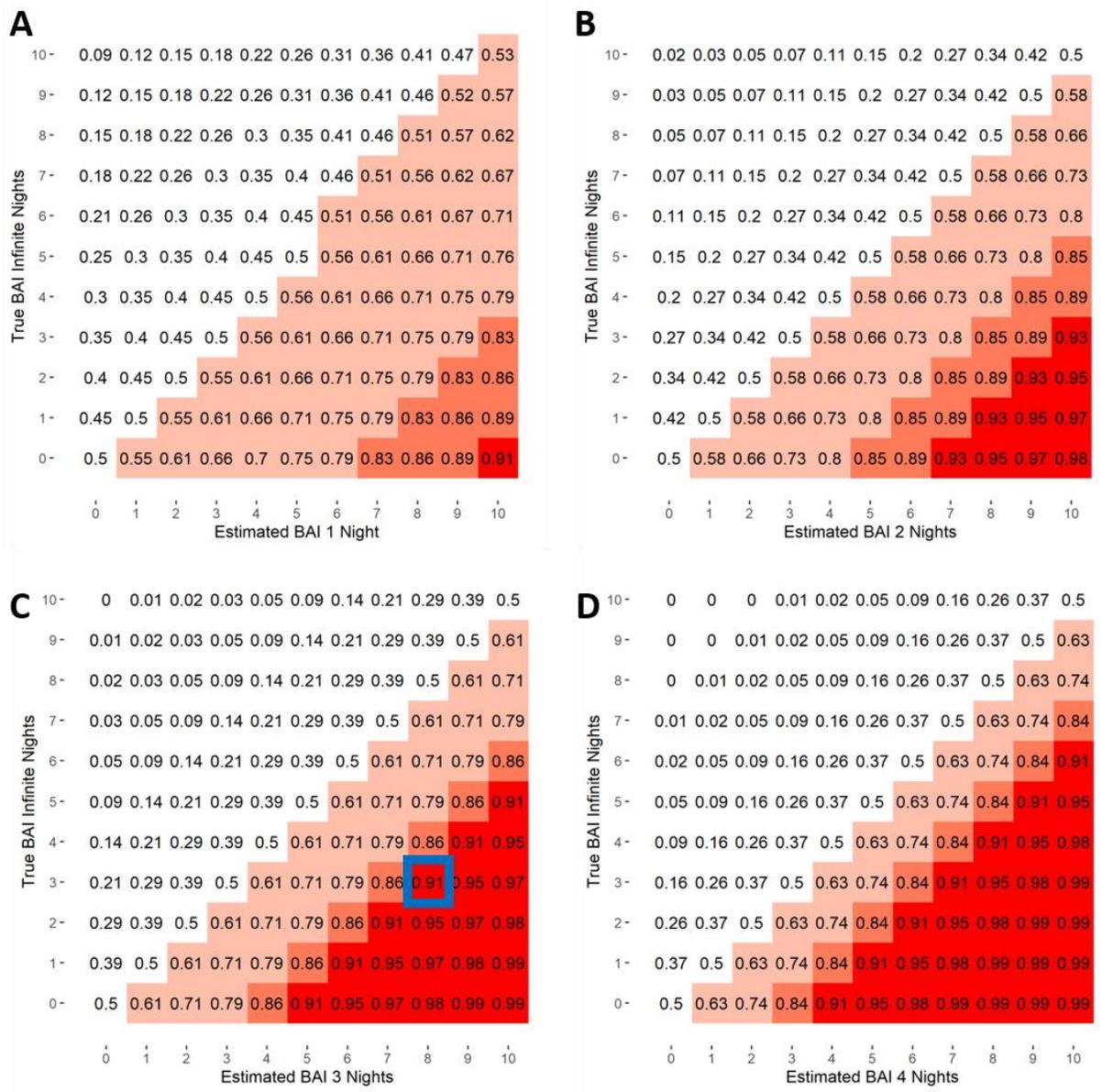

Figure 6- Table of risk probability that a patient's true BAI (based on infinite nights of EEG recordings) is greater than or equal to their estimated BAI if the estimate is calculated from a single night, or when it is averaged from 2, 3 or 4 nights.

*4. Discussion*

Building on the idea that BAI may serve as an accessible biomarker of brain aging, our study quantifies the accuracy of BAI estimates at the individual level, thus providing guidelines for eventually employing such estimates in patient care or clinical trials. We find that the average standard deviation in BAI within individual patients (SD) is approximately 7.5 years. Moreover, a simple "signal+noise" model provides a good description of the data, with night-to-night variability predictably decreasing as $SD/\sqrt{n}$ as the number of available overnight sleep EEG measurements increases. In particular, averaging over four consecutive nights (or four nights within a few days of each other) reduces noise in BAI estimates by more than half, to 3.0. At-home sleep monitoring devices, such multi-day sleep measurements are increasingly practical. Thus, these findings bring BAI one step closer to being useable as a measure of brain health in individual patients, whereas previous work has shown results applicable only at the population level.

Interestingly, our estimate of SD = 7.5 for night-to-night BAI variability within individual patients is close to the average standard deviation *between* patients found in a previous brain age study among healthy subjects (SD = 8.5 years) (Sun et al. 2019). These results suggest that in a healthy population (mean BAI of 0), much of the interpatient variability is simply due to night-to-night variability in sleep microstructure.

While the main aim of our present study was to measure the night-to-night variability in patient brain age, it is interesting that the average BAI in our study is 13 years. We believe there is a simple explanation for this. Patients in the EMU are often kept awake late into the night in an attempt to provoke seizures, which may result in a relative paucity of N3 sleep. This lower prevalence of epochs classified as N3 sleep is normal / typical in older people (Edwards et al. 2010; Varga et al. 2016). Indeed, this idea is quantified in the brain age model itself, where the coefficient associated with N3 sleep is negative. As measuring night-to-night variability is the aim of this analysis, a shift in the mean BAI is not expected to alter the results of this study. Another possible contributing factor to the high brain age in our cohort is the high prevalence of epilepsy patients. This explanation would be in agreement with recent studies showing that MRI-based BAI is higher in patients with medically refractory epilepsy as compared to non-epileptic patients (Pardoe et al. 2017), and another study showing that MRI-based BAI was higher in five out of six epileptic categories (all but extra-temporal lobe focal epilepsy) when compared to non-epileptic control patients (Sone et al. 2019). Nevertheless, while BAI was higher on average for epileptic patients vs. those without an established epilepsy diagnosis, in our data this difference was not statistically significant. The relatively small sample size is a possible explanation.

Night-to-night stability of the BAI is in keeping with other estimates of night-to-night variability/stability of sleep physiology. These include EEG patterns using high-density EEG (Massimini et al. 2004), sleep EEG data from twins (Hori 1986; Gorgoni et al. 2019), heart rate response to arousal from sleep (Gao et al. 2017), performance deficits following sleep deprivation (Kuna et al. 2012), and sleep stability assessed by cardiopulmonary coupling (Thomas et al. 2018).

Our present results provide guidance for how to interpret brain age index values measured for an individual in light of prior population studies. Sun et al. previously showed the population-level difference between patients with diabetes and those without is 3.5 years, and patients with neurological disorders had a BAI that was an average of 4.0 years higher than that of healthy control patients (Sun et al. 2019). Combined with our current results, we see that an individual with a four-night (average) BAI of 8 years or greater has a 90% chance that their true BAI is at least 4, raising the probability that this patient indeed has underlying brain pathology. This information may serve as a warning signal of the possible presence of neurodegenerative disease or other chronic diseases that may have secondary effects on brain function, warranting further investigation or monitoring.

There are several limitations in our study. (1) Our cohort is comprised of EMU patients. While EEG epochs with epileptiform activity were excluded from the study, 73% of our patient cohort had a clinical diagnosis of epilepsy. Furthermore, out-of-home sleep is typically thought to be disrupted, particularly during the first night. Therefore, this patient cohort may not represent the general patient population. (2) Our analysis assumes a uniform pre-test probability distribution for the brain age index. However, since the brain age distribution is unknown population-wide, this assumption seems reasonable, as it avoids assuming healthy or sick BAI values. This assumption also supports the overarching goal of the algorithm as an accurate biomarker for any patient, whether healthy or sick, without assuming the patient's state beforehand. Nevertheless, future research should investigate whether it is useful to create more specific pretest probability distributions for specific populations, e.g., based on gender, race, medical symptoms, or other clinical characteristics. (3) Sleep staging was performed algorithmically as opposed to being hand-labeled by experts. However, this may be argued as a strength, since it reduces between-expert variation in sleep staging as an additional potential source of variation. (4) 61% of patients received antiepileptic drugs during their stay at the EMU which may have affected the EEG signal and consequently the BAI calculation.

*5. Conclusion*

In conclusion, averaging brain ages calculated from multiple nights reduces the effect of night-to-night variability, providing a biomarker of brain health interpretable at the level of an individual. We have provided figures that can serve as clinical guides in applying EEG-based brain age as a biomarker of brain aging.

**Conflict of Interest Statement**

**Acknowledgement**

This research was conducted while Dr. Westover was a Breakthroughs in Gerontology Grant recipient, supported by the Glenn Foundation for Medical Research and the American Federation for Aging Research; through the American Academy of Sleep Medicine through an AASM Foundation Strategic

Research Award; and by grants from the NIH (1R01NS102190, 1R01NS102574, 1R01NS107291, 1RF1AG064312).**References**

Brunner DP, Dijk DJ, Borbely AA. Repeated partial sleep deprivation progressively changes the EEG during sleep and wakefulness. Sleep. 1993;16(2):100–13.

Carne RP, Vogrin S, Litewka L, Cook MJ. Cerebral cortex: An MRI-based study of volume and variance with age and sex. J Clin Neurosci. 2006 Jan;13(1):60–72.

Edwards BA, O'Driscoll DM, Ali A, Jordan AS, Trinder J, Malhotra A. Aging and sleep: Physiology and pathophysiology. Vol. 31, Seminars in Respiratory and Critical Care Medicine. 2010. p. 618–33.

Gao X, Azarbarzin A, Keenan BT, Ostrowski M, Pack FM, Staley B, et al. Heritability of heart rate response to arousals in twins. Sleep. 2017 Jun 1;40(6).

Gorgoni M, Reda F, D'Atri A, Scarpelli S, Ferrara M, De Gennaro L. The heritability of the human K-complex: A twin study. Sleep. 2019;42(6).

Gramfort A, Luessi M, Larson E, Engemann DA, Strohmeier D, Brodbeck C, et al. MEG and EEG data analysis with MNE-Python. Front Neurosci. 2013;(7 DEC).

Gramfort A, Luessi M, Larson E, Engemann DA, Strohmeier D, Brodbeck C, et al. MNE software for processing MEG and EEG data. Neuroimage. 2014 Feb 1;86:446–60.

Hori A. Sleep Characteristics in Twins. Psychiatry Clin Neurosci. 1986;40(1):35–46.

Kuna ST, Maislin G, Pack FM, Staley B, Hachadoorian R, Coccaro EF, et al. Heritability of Performance Deficit Accumulation During Acute Sleep Deprivation in Twins. Sleep. 2012 Sep 1;

Larson E. mne-python/_peak_finder.py at maint/0.19 · mne-tools/mne-python · GitHub [Internet]. 2019a [cited 2019 Nov 26]. Available from: https://github.com/mne-tools/mne-python/blob/maint/0.19/mne/preprocessing/_peak_finder.py#L14

Larson E. mne-python/eog.py at maint/0.19 · mne-tools/mne-python · GitHub [Internet]. 2019b [cited 2019 Nov 26]. Available from: https://github.com/mne-tools/mne-python/blob/maint/0.19/mne/preprocessing/eog.py#L89

Lezak MD. Neuropsychological assessment. Oxford University Press; 2012.

Massimini M, Huber R, Ferrarelli F, Hill S, Tononi G. The sleep slow oscillation as a traveling wave. J Neurosci. 2004 Aug 4;24(31):6862–70.

Pardoe HR, Cole JH, Blackmon K, Thesen T, Kuzniecky R. Structural brain changes in medically refractory focal epilepsy resemble premature brain aging. Epilepsy Res. 2017 Jul 1;133:28–32.

Sone D, Beheshti I, Maikusa N, Ota M, Kimura Y, Sato N, et al. Neuroimaging-based brain-age prediction in diverse forms of epilepsy: a signature of psychosis and beyond. Mol Psychiatry. 2019;

Sun H, Jia J, Goparaju B, Huang G Bin, Sourina O, Bianchi MT, et al. Large-scale automated sleep staging. Sleep. 2017 Oct 1;40(10).
17

**Supplementary Tables**

Table S1. The percent of time spent in each stage was used to predict standard deviation in BAI for each patient.

| % Time | Univariate Regression p-value | Multivariate Regression p-value |
|---|---|---|
| N1 | 0.23 | 0.31 |
| N2 | 0.61 | 0.85 |
| N3 | 0.8 | 0.57 |
| REM | 0.14 | 0.16 |
| Awake | 0.52 | 0.81 |

Table S2. The standard deviation in percent of time spent in each stage was used to predict standard deviation in BAI for each patient.

| SD % Time | Univariate Regression p-value | Multivariate Regression p-value |
|---|---|---|
| N1 | 0.9 | 0.96 |
| N2 | 0.23 | 0.22 |
| N3 | 0.54 | 0.19 |
| REM | 0.15 | 0.15 |
| Awake | 0.18 | 0.5 |